\documentclass[12pt,preprint]{aastex}

\newcommand{\bdv}[1]{\mbox{\boldmath$#1$}}

\def\au{{\rm AU}} 
\def\rel{{\rm rel}}
\def\e{{\rm E}}
\def\bpi{{\bdv\pi}}
\def\bmu{{\bdv\mu}}
\def\p{{\partial}}
\def\eps{{\epsilon}}
\def\epar{{{\rm E},\parallel}}
\def\eper{{{\rm E},\perp}}

\begin{document}
\title{Geosynchronous Microlens Parallaxes}

\author{Andrew Gould}
\affil{Department of Astronomy, Ohio State University,
140 W.\ 18th Ave., Columbus, OH 43210, USA; 
gould@astronomy.ohio-state.edu}

\begin{abstract}
I show that for a substantial fraction of planets detected in
a space-based survey, it would be possible to measure the planet
and host masses and distances, if the survey satellite were placed
in geosynchronous orbit. Such an orbit would enable measurement
of the microlens parallax $\pi_\e$ for events with moderately low
impact parameters, $\beta\la 0.05$, which encompass a disproportionate
share of planetary detections.  Most planetary events
yield a measurement of the angular Einstein radius $\theta_\e$.
Since the host mass is given by $M=\theta_\e/\kappa\pi_\e$ where
$\kappa$ is a constant, parallax measurements are the crucial
missing link. I present simple analytic formulae that enable
quick error estimates for observatories in circular orbits
of arbitrary period and semi-major axis, and arbitrary orientation
relative to the line of sight.  The method requires photometric
stability of $\sim 10^{-4}$ on $\sim 1$ orbit timescales.  I show that
the satellite data themselves can provide a rigorous test of 
the accuracy of the parallax measurements in the face
of unknown systematics and stellar variability, even at this 
extreme level.

\end{abstract}


\section{{Introduction}
\label{sec:intro}}

Microlens planet detections routinely yield the planet-star mass
$q=m/M$, but since the host star is generally very difficult to observe,
the host mass $M$ (and so the planet mass $m$) and the system distance
$D_L$ usually remain undetermined.  In principle, one can determine
these quantities from the microlensing event itself, provided the
angular Einstein radius $\theta_\e$ and the microlens parallax $\pi_\e$
are both measured \citep{gould92}
\begin{equation}
M = {\theta_\e\over \kappa \pi_\e},
\quad
\pi_\rel = \theta_\e\pi_\e,
\quad
\kappa\equiv {4 G\over c^2 {\rm AU}} = 8.1 {{\rm mas}\over M_\odot},
\label{eqn:massdis}
\end{equation}
where $\pi_\e = \au/\tilde r_\e$ is the ratio of the 
Earth's orbit to the Einstein radius projected on the observer plane, and
$\pi_\rel\equiv {\rm AU}[D_L^{-1}-D_S^{-1}]$ is the lens-source
relative parallax, with $D_S$ generally being known from direct observation
of the source.  

To date, $\theta_\e$ has been measured in the majority of planetary
events, even though it is rarely ($\sim 0.1\%$) measured in ordinary
events.  This is because planets are detected when the source passes
near or over a caustic induced by the planet, which yields a measurement
of the source size relative to the Einstein radius.

Hence, if a method could be found to measure $\pi_\e$ for a large fraction
of events, the information returned from microlensing planet searches
would be radically improved.  The microlens parallax is actually
a vector $\bpi_\e$, with the direction being that of the lens-source
relative motion.  It can in principle be measured either from the
ground \citep{gould92}, by combining space-based and ground-based
observations \citep{refsdal66}, or from space alone \citep{honma99}.

When I proposed the pure ground-based method \citep{gould92},
I illustrated it with an event that lasted 4 years, so that the
lightcurve oscillated annually around the standard microlensing form.
In practice, such 4-year events are never seen, but $\bpi_\e$ can
be measured from less regular features in shorter events.  However,
typical events have timescales $t_\e\sim 20\,$days, so that
the great majority are too short to permit a measurement.
\citet{honma99} proposed a purely space-based method using 
{\it Hubble Space Telescope (HST)} observations.  Since the {\it HST}
orbital period is only 1.5 hours, the oscillations are obviously
much shorter than $t_\e$.  However, the {\it HST} orbital radius is
only $R_\oplus$, whereas typically $\tilde r_\e \sim 1-10\,\au$.
Hence, the oscillations would typically be minuscule.  To overcome this
obstacle, \citet{honma99} proposed that the observations should be
made during caustic crossings, which he showed dramatically increase
the strength of the signal.  However, alerting {\it HST} to imminent
caustic crossings is extremely difficult, more so for planetary events,
and in fact such measurements have never been carried out, nor even
attempted.  The ground-space combination originally proposed by
\citet{refsdal66} remains feasible in principle, but each event
must be observed very frequently \citep{gaudi97}, implying a huge
space-craft investment.  Recently, \citet{gould12} showed that
a much simpler and cheaper ground-space approach was feasible for
high-magnification events, but these constitute a small subset.

Here I show that a single survey satellite in geosynchronous orbit 
is a near-optimal solution to this problem.  As in my original
proposal, oscillations are induced on a standard microlensing
lightcurve,  but since these have periods of one day rather than one
year, they are shorter than the event.  The amplitude of the oscillations
is 6.6 times larger than those of the \citet{honma99} proposed {\it HST}
observations.  Moreover, no special warning system is required.  It
is true that this approach only works for moderately high-magnification
($A\ga 10$) events, but these are more sensitive to planets than
more typical events \citep{gouldloeb}.  Hence, this approach can
yield parallaxes for a large fraction of planetary events.  Moreover,
such a geosynchronous survey satellite is a very real possibility.
The \citet{nwnh} report recommended a 
{\it Wide Field Infrared Space Telescope (WFIRST)} as its highest
priority.  A large fraction of the observing time would be devoted
to microlensing planet searches.  Although originally recommended for
L2 orbit, a geosynchronous orbit is now being actively considered.
Hence, it is most timely to investigate the implications of
such an orbit for microlensing parallaxes.

\section{{Analytic Treatment of Single-Observatory Parallax}
\label{sec:satpar}}

Consider an observatory in circular orbit with period $P$ and
semi-major axis $a$.  If this is an Earth orbit, these two quantities
are obviously related, but here I begin with the general case.  Let
the microlensing target be at a latitude $\lambda$ with respect to
the plane of this orbit, so that the projected semi-minor axis is
$b=a\sin\lambda$. I then normalize these axes to an AU
\begin{equation}
\eps_\parallel \equiv \eps = {a\over\au};
\quad
\eps_\perp = {a\sin\lambda\over\au};
\label{eqn:epseq}
\end{equation}

Now consider that continuous observations of a microlensing event
are made at a rate $N$ per Einstein timescale $t_\e$, with signal-limited
flux errors
\begin{equation}
\sigma^2 = \sigma_0^2 A.
\label{eqn:erroreq}
\end{equation}
This condition is not as restrictive as may first appear.  We are only
concerned with scaling of the errors when the source is relatively
highly magnified, so that this form need not hold all the way down
to baseline, where sky and blending may be important. 
Let the lens pass the source on its right at an angle
$\theta$ relative to the major projected axis of the satellite orbit,
with impact parameter $\beta$ (in units of $\theta_\e$).  Then the
lens-source separation vector $\bf u$ (in units of $\theta_\e$) 
is given by 
\begin{equation}
{\bf u} = 
(\tau\cos\theta - \beta\sin\theta + \eps_\parallel\pi_\e\cos(\omega t+\phi),
\tau\sin\theta + \beta\cos\theta + \eps_\perp\pi_\e\sin(\omega t+\phi))
\label{eqn:ueval}
\end{equation}
where $\tau\equiv (t-t_0)/t_\e$, $t_0$ is the time of closest approach,
$\omega=2\pi/P$, and $\phi$ is the orbit phase relative to peak.
By definition, $\bpi_\e = (\pi_\epar,\pi_\eper)= \pi_\e(\cos\theta,\sin\theta)$.
Hence,
\begin{equation}
{\p u\over \p \pi_\epar} =
{\eps_\parallel\tau\cos(\omega t+\phi) + \eps_\perp\beta\sin(\omega t+\phi)\over u},
\qquad
{\p u\over \p \pi_\eper} = 
{-\eps_\parallel\beta\cos(\omega t+\phi) + \eps_\perp\tau\sin(\omega t+\phi)\over u}
\label{eqn:upartials}
\end{equation}

\begin{figure}
\plotone{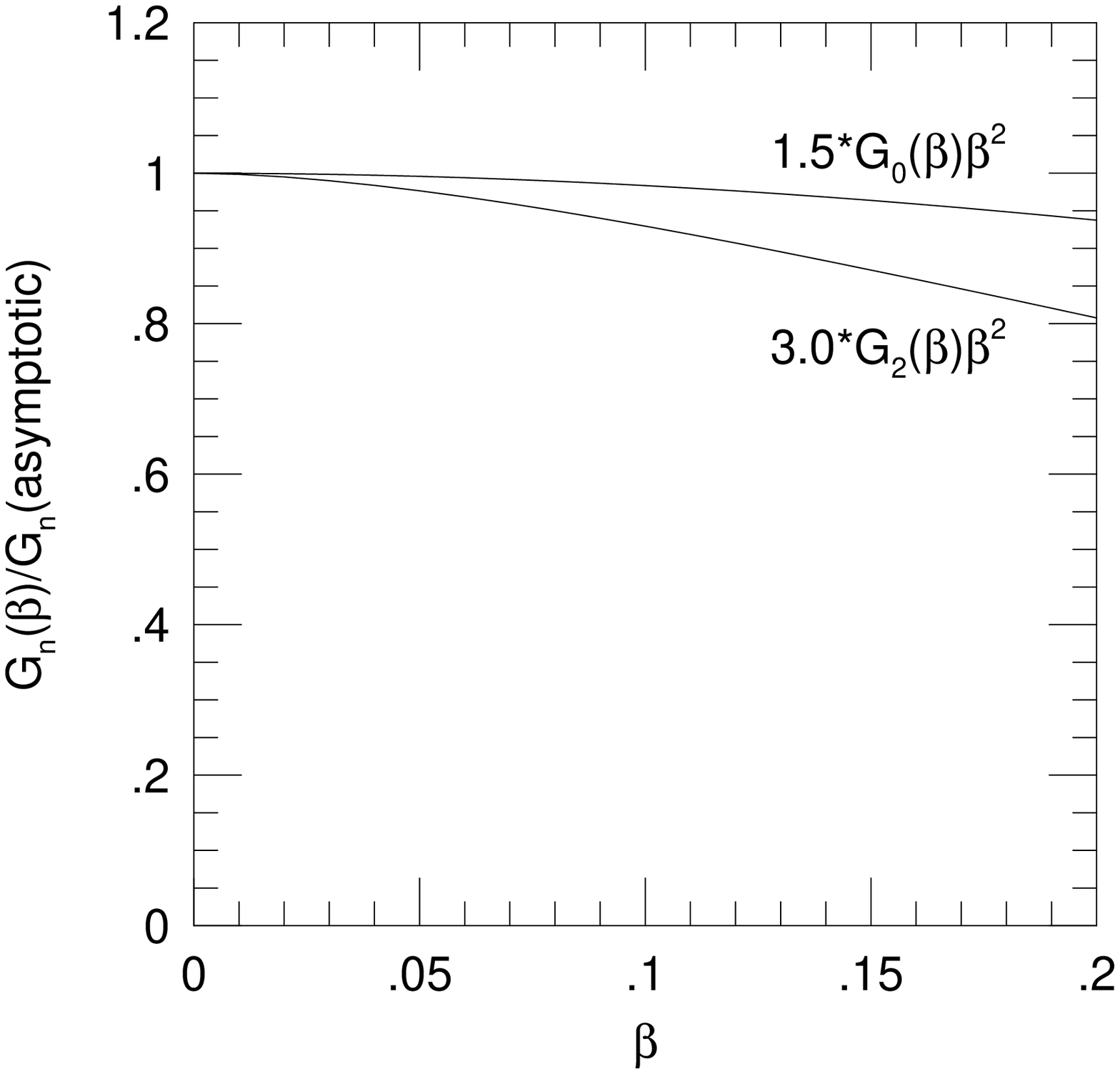}
\caption{\label{fig:deviate}
Functional forms of $G_0(\beta)$ and $G_2(\beta)$ relative to the
limiting forms given by Equation~(\ref{eqn:geval}).  These approximations
are excellent for $\beta<0.05$ and very good for $\beta<0.1$.
}
\end{figure}

The magnification is given by \citep{einstein36}
\begin{equation}
A = {u^2+2\over u\sqrt{u^2+4}},
\qquad
{\p \ln A\over \p u} = -{8\over u(u^2+2)(u^2+4)},
\label{eqn:mageq}
\end{equation}
so that the derivatives of the flux with respect to the parallax parameters are
\begin{equation}
{\p F\over \p\pi_\epar} = F_s{\p \ln A\over \p u} A {\p u\over \p \pi_\epar},
\qquad
{\p F\over \p\pi_\eper} = F_s{\p \ln A\over \p u} A {\p u\over \p \pi_\eper} ,
\label{eqn:fpartials}
\end{equation}
where $F_s$ is the unmagnified source flux.
In order to make a Fisher-matrix analysis,
I first assume that the magnification does not change much over an
orbit. This will ultimately restrict the result to events for which
$t_{\rm eff}\equiv \beta t_\e > P$. I will discuss the more general
case further below. I then evaluate the mean contribution to
the Fisher matrix, averaged over one orbit
\begin{equation}
{1\over \sigma^2}\langle\biggl({\p F\over \p\pi_\epar}\biggr)^2\rangle
= {F_s^2\over\sigma_0^2}\langle\biggl({\p u\over \p\pi_\epar}\biggr)^2\rangle
A\biggl({\p \ln A\over \p u}\biggr)^2
= {F_s^2\over 2\sigma_0^2}{64(\eps_\parallel^2\tau^2 + \eps_\perp^2\beta^2)
\over u^5(u^2+4)^{5/2}(u^2+2)}
\label{eqn:parpar}
\end{equation}
\begin{equation}
{1\over \sigma^2}\langle\biggl({\p F\over \p\pi_\eper}\biggr)^2\rangle
= {F_s^2\over 2\sigma_0^2}{64(\eps_\perp^2\tau^2 + \eps_\parallel^2\beta^2)
\over u^5(u^2+4)^{5/2}(u^2+2)}
\label{eqn:perper}
\end{equation}
\begin{equation}
{1\over \sigma^2}\langle{\p F\over \p\pi_\epar}{\p F\over \p\pi_\eper}\rangle
= {F_s^2\over 2\sigma_0^2}{64(\eps_\perp^2-\eps_\parallel^2)\tau\beta
\over u^5(u^2+4)^{5/2}(u^2+2)}
\label{eqn:parper}
\end{equation}
and so obtain the inverse covariance matrix by integrating
over all observations
\begin{equation}
B = {N F_s^2\over \sigma_0^2}
\left(\matrix{\eps_\parallel^2 G_2 + \eps_\perp^2 G_0 & 0\cr
0 & \eps_\perp^2 G_2 + \eps_\parallel^2 G_0\cr}\right)
\label{eqn:bmat}
\end{equation}
where
\begin{equation}
G_n(\beta) \equiv \int_0^\infty dx{64\,x^n \beta^{2-n}
\over (x^2+\beta^2)^{5/2}(x^2+\beta^2+4)^{5/2}(x^2+\beta^2 + 2)}.
\label{eqn:gdef}
\end{equation}
In the limit $\beta\ll 1$, it is straightforward to show that
\begin{equation}
G_0(\beta)\rightarrow {2\over 3}\beta^{-2},
\quad
G_2(\beta)\rightarrow {1\over 3}\beta^{-2}.
\label{eqn:geval}
\end{equation}
Figure~\ref{fig:deviate} shows that this approximation holds quite
well for $\beta\la 0.1$ (i.e., $A_{\rm max}\ga 10$). Using this approximation,
we obtain,
\begin{equation}
\left(\matrix{\sigma(\pi_\epar) \cr \sigma(\pi_\eper)}\right)
= \sqrt{3\over N}{\sigma_0\over F_s}{\beta\over \epsilon}
\left(\matrix{(1+2\sin^2\lambda)^{-1/2}\cr (2+\sin^2\lambda)^{-1/2}}\right)
\label{eqn:stdev}
\end{equation}
Since the ratio of the quantities in the right-hand vector is never greater
than $\sqrt{2}$ (and is $\sqrt{3/2}$ for $\lambda=30^\circ$, which is typical
of Galactic bulge fields and an equatorial orbit), the parallax error
is reasonably well characterized by the harmonic rms of these two values,
\begin{equation}
\sigma(\pi_\e)\simeq\biggl({B_{11} + B_{22}\over 2}\biggr)^{-1/2}
= 0.023
\biggl({N\over 10000}\bigg)^{-1/2}
\biggl({1+\sin^2\lambda\over 1.25}\bigg)^{-1/2}
\biggl({\sigma_0/F_s\over 0.01}\bigg)
\biggl({\beta\over 0.05}\bigg)
\biggl({\eps\over 6.6\,R_\oplus/\au}\bigg)^{-1}
\label{eqn:sigmapie}
\end{equation}

\section{{Application to {\it WFIRST}}
\label{sec:wfirst}}

The fiducial values in Equation~(\ref{eqn:sigmapie}) are plausible for
a {\it WFIRST} style mission.  For example, a continuous sequence
of 3-minute exposures yields $N\sim 10000$ for a $t_\e=20\,$day event.
A large fraction of sources would have 1\% errors in such an exposure,
whether continuous or co-added.  Note that the formula does not really
specify $\sigma_0$ and $N$ separately, but just $\sigma_0/\sqrt{N}$.
Moreover, for typical $t_\e=20\,$day events and a $P=1\,$day orbit,
the assumptions of the derivation basically apply for $\beta=0.05$,
i.e., $t_{\rm eff} = \beta t_\e = P$. Of course, the integral in
Equation~(\ref{eqn:gdef}) goes to infinity, while the observations do not,
but the overwhelming contribution to this integral is from a few
$t_{\rm eff}$ near peak.  The fiducial parallax error in 
Equation~(\ref{eqn:sigmapie}) is quite adequate to measure the mass
and distance of disk lenses but somewhat marginal for bulge lenses.
For example, for $M=0.5\,M_\odot$, a typical disk lens at $D_L = 4\,$kpc
has $\pi_\e\sim 0.18$, but a typical bulge lens at $D_S - D_L = 0.75\,$kpc
has $\pi_\e\sim 0.06$.  At the very least, however, such a measurement
would distinguish between bulge and disk lenses.

\section{{Systematics and Stellar Variability}
\label{sec:var}}

Naively, Equation~(\ref{eqn:sigmapie}) appears to require ``effective
errors'' of $N^{-1/2}\sigma_0/F_s\sim 10^{-4}$, which must be achieved
in the face of both astrophysical and instrumental effects.  In particular,
it would seem to challenge
the systematics limit for real crowded-field photometry, even from space.
However, what is actually required is not control of systematics to
this level, but only control of systematic effects that correlate
with orbital phase.  This requirement is still not trivial for geosynchronous
orbit because of phase-dependent temperature variations, but it
is not as intractable as vetting all systematic errors at this level.

Nothing is known about stellar variability at this level in $H$ 
(likely WFIRST passband).  Even {\it Kepler} has only a handful of
stars for which it can make measurements of this precision
in its optical passband
(Fig.~4 of \citealt{mcquillan12}).  However, again what is relevant
is not variability per se, but variability on 1-day timescales,
and \citet{mcquillan12} report that FGK stars typically vary on
$>5$-day timescales.  Moreover, most forms of intrinsic variability
are very subdued in $H$ band relative to the optical.  See \citet{mb10523}
for a spectacular example of this in a microlensing event.  Finally,
\citet{mcquillan12} find that variability declines with increasing
proper motion (so presumably age), and the bulge is mostly much
older than the disk, though perhaps not entirely \citep{bensby13}

\section{{Rigorous Test of Parallax Accuracy}
\label{sec:test}}

\begin{figure}
\plotone{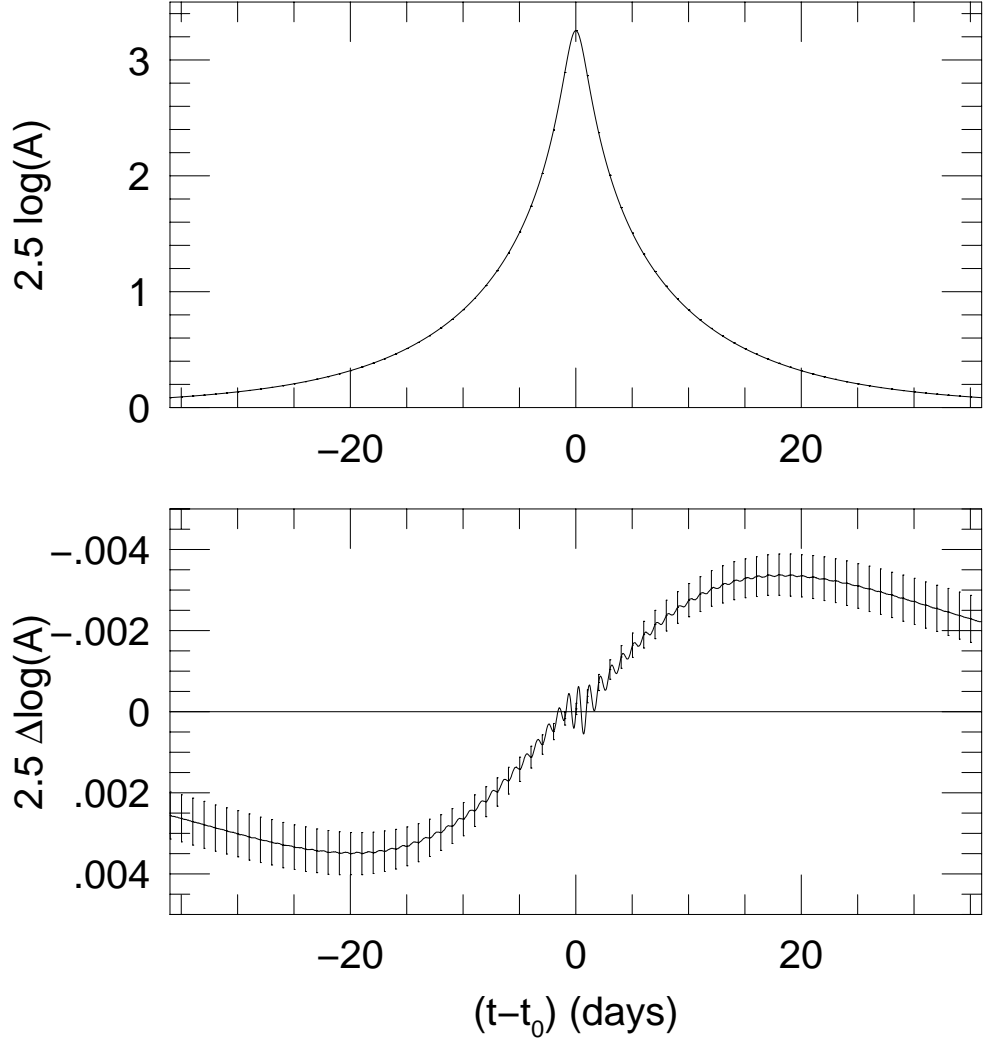}
\caption{\label{fig:lc}
Simulated lightcurve of event with $t_\e=20\,$days, $\beta=0.05$, $t_0$
at the Vernal Equinox, and
parallax $\pi_{\e,N}=\pi_{\e,E}=0.1$, observed from a geosynchronous
equatorial orbit.  The error bars are binned by day for display but
the observations are assumed many times per day.
}
\end{figure}

\begin{figure}
\plotone{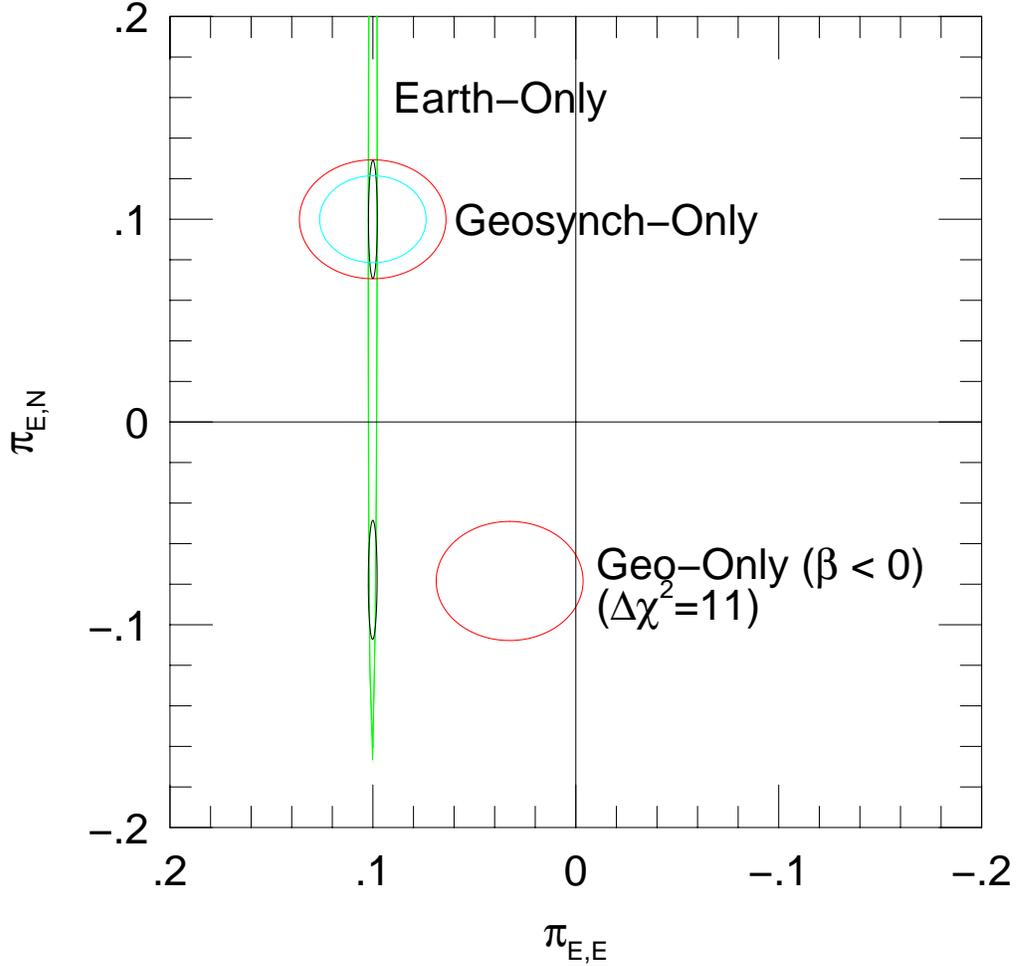}
\caption{\label{fig:ell}
Error ellipses ($1\,\sigma$) for the event shown in Figure~\ref{fig:lc}
using Earth-orbit-only (green), geosynchronous-only (red) and combined
(black) information.  The lightcurve asymmetry (Fig.~\ref{fig:lc})
due to the Earth's orbit yields only 1-D parallax information, but this
serves as a critical check on the accuracy of the geosynchronous-orbit
parallax.  The secondary minimum at $\pi_{\e,N}\sim -0.07$ is due to
$(\beta\rightarrow -\beta)$ degeneracy.  In this example, it is 
disfavored by $\Delta\chi^2=11$ based on geosynchronous data and by 15
based on all data.  The analytic error estimate (based on infinite data)
is shown in cyan.
}
\end{figure}

One would nevertheless like
a rigorous test that undetectable variability and systematic effects
are not corrupting the results.  Fortunately the satellite data
themselves provide such a test.  \citet{gmb94} showed that
even very short microlensing events yield ``one-dimensional parallaxes''
from the asymmetry in the lightcurve induced by the Earth's
(approximately constant) acceleration toward the Sun.  That is,
$\pi_{\e,\odot,\parallel}$ is well constrained while 
$\pi_{\e,\odot,\perp}$ is very poorly constrained.  Normally,
such 1-D parallaxes are not considered useful because the amplitude of $\pi_\e$
(and so $M$) is also poorly constrained.  However, in the present case
these 1-D parallaxes serve two very important functions.
Figure~\ref{fig:lc} shows the
lightcurve of a simulated event with 72 days of observations centered
on the Vernal Equinox in each of 5 years, similar to the anticipated
schedule of WFIRST.  The event is assumed to have the fiducial
parameters from Equation~(\ref{eqn:sigmapie}): $t_\e=20\,$days, $\beta=0.05$,
$N^{-1/2}(\sigma_0/F_s)=10^{-4}$, with $t_0$ right at the Vernal Equinox.
The lower panel shows the difference between this event as observed
from a geosynchronous orbit, and the same event observed from an
inertial platform.  Note that since the orbit is equatorial,
$(\pi_{\e,\parallel},\pi_{\e,\perp})=(\pi_{\e,E},\pi_{\e,N})$,
i.e., the components of $\bpi_\e$ in the East and North directions
projected on the sky.  The event is at $(\alpha,\delta)=$(18:00:00,-30:00:00),
and is assumed  to have a parallax $(\pi_{\e,E},\pi_{\e,N})=(0.1,0.1)$.
The oscillations near peak are due to the
satellite's orbital motion, while the asymmetries in the wings are 
from the Earth's orbit.  The latter are much larger, but yield
only 1-D information.  This is illustrated in Figure~\ref{fig:ell},
which shows the error ellipses due to 1) Earth-only, 2) satellite-only,
3) Earth+satellite.  Also shown is the result of the analytic calculation
(which assumed observations extending to infinity).  Note that there
are two sets of solutions in each case, corresponding to the 
($\beta\rightarrow -\beta$) degeneracy \citep{smith03,gould04}.

The first point is that if $\pi_{\e,E}$ as derived from geosynchronous
parallax agrees with the much more precise value
derived from the Earth's orbit, then one can have good
confidence in $\pi_{\e,N}$, for which there
is no direct test.  This is true on an event by event basis, but
more true for the ensemble of parallax measurements.  

The second point is that these two parallax measurements will be automatically
combined in the fit to any lightcurve, which means that $\pi_{\e,E}$
will be much better determined than $\pi_{\e,N}$.   Depending on the
relative values of these two components, this may add important
information in some cases.

Finally, the ensemble of geosynchronous measurements of $\pi_{\e,E}$
provides a test of the accuracy of the 1-D Earth-orbit measurement
of this quantity from the wings of the event.  This is important because
the source stars may show variability on 10-day timescales, even if they
do not vary on 1-day timescales.

Using techniques similar to those used to derive Equation~(\ref{eqn:stdev}),
it is straightforward to show that the 1-D parallax error due to 
Earth acceleration for short, relatively high-magnification
events (and infinite observations) is
\begin{equation}
\sigma(\pi_{\e,\odot,\parallel}) = \sqrt{3\over N}
{\sigma_0\over F_s}
\biggl({t_\e\over 58\,{\rm day}}\biggr)^{-2}\eta^{-1},
\label{eqn:earthorbit}
\end{equation}
where 58 days is one radian of the Earth's orbit and $\eta$
is the  projected Earth-Sun separation in AU at time $t_0$.
Since $\eta\sim 1$ for bulge observations made during the
equinoxes, the ratio of geosynchronous-to-Earth parallax errors is
\begin{equation}
{\sigma(\pi_{\e})_{\rm geosynch}\over
\sigma(\pi_{\e,\odot,\parallel})_{\rm Earth}}
\simeq 
20\biggl({t_\e\over 20\,{\rm day}}\biggr)^{2}\,
{\beta\over 0.05} = {t_\e\over 1\,\rm day}\,{\beta t_\e\over 1\,\rm day}.
\label{eqn:compsigmas}
\end{equation}
Since the effective $\beta t_\e\ga P=1\,$day,
this implies that the Earth-orbit parallax will essentially always yield
a precise check on the geosynchronous parallax in one direction.

\section{{Discussion}
\label{sec:discuss}}

The derivation underlying Equation~(\ref{eqn:sigmapie}) breaks down
for $t_{\rm eff}\la P$: the errors continue to decline with falling
$\beta$, but no longer linearly. They also become dependent on the
orientation and phase of the orbit in a much more complicated way.  From
the present perspective, the main point is that the formula with
$\beta\rightarrow P/t_\e$ provides an upper limit on the errors
for events with yet higher peak magnification.

Next, the errors derived here assume a point-lens event. However, since
the observations would be near-continuous, it is likely that 
caustic crossings or near approaches would be captured.  As pointed
out by \citet{honma99} such caustic effects can significantly enhance
the signal.

Another feature of these (and most) parallax measurements is that they
work better at low mass, simply because $\pi_\e\propto M^{-1/2}$ is
bigger.  Space-based microlensing measurements have the potential
to directly detect the lens when it is more massive (so, typically,
brighter).  For example, as the lens and source separate after 
(or before) the event, their joint light becomes extended and
the centroids of the blue and red light separate (if the source
and lens are different colors).  These effects allowed \citet{bennett06}
and \citet{dong09} to measure the host masses in two different planetary
events using followup {\it HST} data.  Because geosynchronous parallax
works better at low mass, while photometric/astrometric methods
work better at high mass, they are complementary.

Finally, I note that such parallaxes would be of great interest
in non-planetary events as well. Without $\theta_\e$, such measurements
do not yield masses and distances, but they do serve as important inputs into
Bayesian estimates of these quantities.  Moreover, since the direction
of $\bpi_\e$ is the same as that of the lens-source relative
proper motion $\bmu_\rel$, a parallax measurement provides an important
constraint when trying to detect/measure the source-lens displacement
away from the event.  If the magnitude $\mu_\rel$ can be measured from
these data, then so can $\theta_\e = \mu_\rel t_\e$, which in turn yields
the mass and distance.

\acknowledgments

This work was supported by NSF grant AST 1103471 
and NASA grant NNX12AB99G.

\end{document}